\documentstyle[twoside,fleqn,espcrc2]{article}

\newcommand{\AmS}{{\protect\the\textfont2
  A\kern-.1667em\lower.5ex\hbox{M}\kern-.125emS}}

\title{Muons from strangelets}

\author{M.Rybczy\'nski, Z.W\l odarczyk
        \address{Institute of Physics, Pedagogical 
        University, Kielce, Poland \\
        emails: mryb@pu.kielce.pl and wlod@pu.kielce.pl}
        and 
        G. Wilk\address{The Andrzej Soltan Institute for Nuclear Studies, 
        Nuclear Theory Department, Warsaw, Poland\\
        email: wilk@fuw.edu.pl}}
       
\begin{document}

\begin{abstract}
The hypothesis is discussed that muon bundles of extremaly high
multipli\-city observed recently by ALEPH detector (in the dedicated
cosmic-ray run) can originate from the strangelets colliding with the
atmosphere. 
\end{abstract}

\maketitle

\section{INTRODUCTION}

In the astrophysical literature \cite{K} one can find a number of
phenomena which can be regarded as a possible manifestation of the
existence of the so called {\it Strange Quark Matter} (SQM) (in the
form of lumps called strangelets), extremaly interesting possibility
of a possible new stable form of matter. They include, among others,
anomalous cosmic ray burst from {\it Cygnus X-3}, extraordinary high
luminosity gamma-ray bursts from the {\it supernova remnant N49} in
the Large Magellanic Cloud or {\it Centauro} type events. There are
also several reports suggesting direct candidates for the SQM. In
particular, anomalous massive particles, which can be interpreted as
strangelets, have been apparently observed in cosmic ray experiments
\cite{K}. All this makes a search for other possible candidates or
signals for SQM extremaly interesting topic.

Proceeding along this line we would like to bring ones attention to
the recent (still unpublished, however) data from the cosmic ray run 
of the ALEPH detector at CERN-LEP experiment. The hypothesis which we
shall discuss in what follows is that, if confirmed, the muon bundles
of extremely high multiplicity observed recently by ALEPH in its
dedicated cosmic-ray run \cite{COSMOLEP} can originate from the
strangelets propagating through the atmosphere and interacting with
the air nuclei. 

\section{MUON BUNDLES FROM CosmoLEP}

Why the CosmoLEP data are potentially so important? The reson is
twofold. First, the studies of high multiplicity cosmic muon events
(called muon bundles) is potentially very important source of
information about the composition of primary cosmic rays. It is
because muons transport (in essentially undisturbed way) significant
information on the first interaction of the cosmic ray particle with
atmosphere. In comparison electromagnetic cascades are more
calorimetric in nature and less sensitive to any model uncertainties,
which could be important for establishing the primary spectrum. The
second point has to do with the fact that multi-muon bundles have
never been studied with such precise detectors as provided by LEP
program at CERN, nor have they been studies at such depth as at CERN
\cite{LEP}. The underground location of the LEP detectors (between
$30$ and $140$ meters) is ideal for the muon based experiments
because the corresponding muon momentum cut-off is then between $15$
and $70$ GeV, i.e., in the most sensitive range from the point of
view of the primary interaction, where interaction and decay
probabilities are equal at the starting point of the cascade.

The present situation is following. Data archives from the ALEPH runs
have revealed a substantial collection of cosmic ray muon events.
More than $3.7\cdot 10^5$ muon events have been recorded in the
effective run time $10^6$ seconds. Multi-muon events observed in the
$16~{\rm m}^2$ time-projection chamber with momentum cut-off $70$ GeV
have been analysed and good agreement with the Monte Carlo
simulations (performed using {\it Corsika} code \cite{COR}) obtained for
multiplicities $N_{\mu}$ between $2$ and $40$. However, there are $5$
events with unexpectedly large multiplicities $N_{\mu}$ (up to $150$)
which rate cannot be explained, even assuming pure iron primaries.
They will be our central point of interest here.

\section{SOME FEATURES OF STRAN\-GE\-LETS}

For completeness we shall summarize now features of strangelets and
their propagation through the atmosphere, which will be relevant to
our further discussion. The more detailed information can be found in
\cite{WW}. Typical SQM consists of roughly equal number of up ($u$),
down ($d$) and strange ($s$) quarks and it has been argued to be the
true ground state of QCD \cite{SQM,SQM1}. For example, it is
absolutely stable at high mass number $A$ (excluding weak interaction
decays of strange quarks, of course) and it would be more stable than
the most tightly bound nucleus as iron (because the energy per barion
in SQM could be smaller than that in ordinary nuclear matter). On the
other hand it becomes unstable below some critical mass number
$A_{crit}$, which is of the order of $A_{crit} = 300 - 400$, depending
on the various choices of relevant parameters \cite{SQM1}. At this
value of $A$ the separation energy, i.e., the energy which is
required to remove a single barion from a strangelets starts to be
negative and strangelet decays rapidly by evaporating neutrons.

In \cite{WW} we have demonstrated that the geometrical radii of
strangelets $R=r_0 A^{1/3}$ are comparable to those of ordinary
nuclei of the corresponding mass number $A$ (i.e., in both cases
$r_0$ are essentially the same). We have shown at the same place how
it is possible that such big objects can apparently propagate very
deep into atmosphere. The scenario proposed and tested in \cite{WW}
was that after each collision with the atmosphere nucleus strangelet
of mass number $A_0$ becomes a new one with mass number approximately
equal $A_0-A_{air}$ and this procedure continues unless either
strangelet reaches Earth or (most probably) disintegrates at some
depth $h$ of atmosphere reaching $A(h)=A_{crit}$.

This results, in a first approximation (in which $A_{air} << A_{crit}
< A_0$), in the total penetration depth of the order of
\begin{equation}
\Lambda\, \simeq\, \frac{4}{3}\, \lambda_{N-air}\,
\left(\frac{A_0}{A_{air}}\right)^{1/3} \label{eq:L}
\end{equation}
where $\lambda_{N-air}$ is the usual mean free path of the nucleon in
the atmosphere.

\section{RESULTS}

\begin{figure}[h]
\setlength{\unitlength}{1cm}
\begin{picture}(7.28,7.11)
\includegraphics{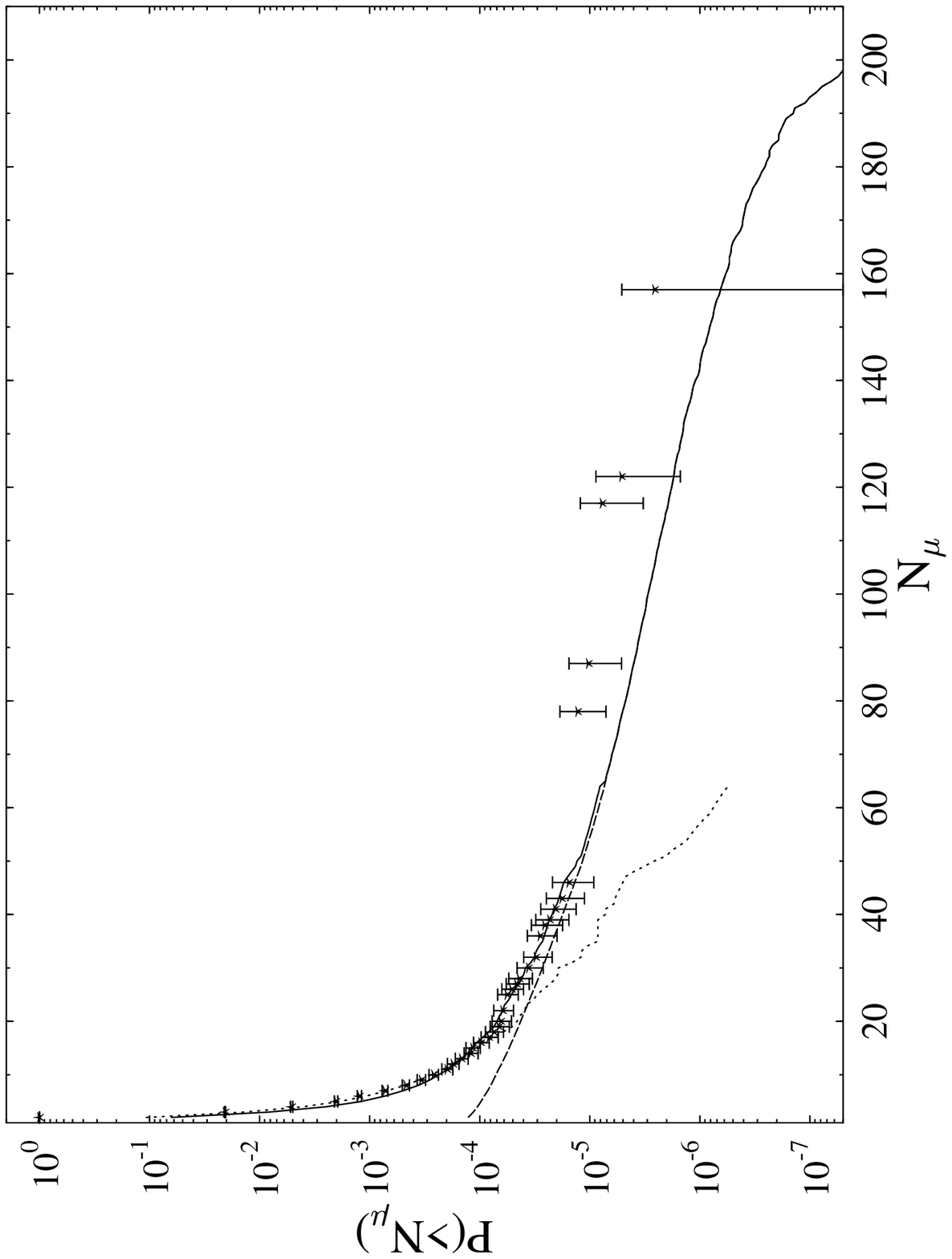}
\end{picture}
\vspace{-0.5cm}
\end{figure}
\vspace{-0.5cm}
\begin{minipage}[h]{6.8cm}
\noindent
Fig.1 Integral multiplicity distribution of muons for the CosmoLEP
data (stars), Monte Carlo simulations for primary nuclei with
"normal" composition (dotted line) and for primary strangelets with
$A=400$ (broken line). Full line shows the summary (calculated) distribution.
\end{minipage}
\vspace{0.5cm}

This is the picture we shall use to estimate the production of muon
bundles produced as result of interaction of strangelets with
atmospheric nuclei. We use for this purpose the
SHOWERSIM \cite{SHOWER} modular software system specifically modified
for our present purpose. Monte Carlo program describes the
interaction of the primary particles at the top of atmosphere and
follows the resulting electromagnetic and hadronic cascades through
the atmosphere down to the observation level. Muons with momenta
exceeding $70$ GeV are then registered in the sensitive area of
$16~{\rm m}^2$ (randomly scattered in respect to the shower axes).
Primaries initiated showers were sampled from the usual power
spectrum $P(E) \propto E^{-\gamma}$ with the slope index equal to
$\gamma = 2.7$ and with energies above $10\cdot A$ TeV. 

\begin{figure}[h]
\setlength{\unitlength}{1cm}
\begin{picture}(7.28,7.11)
\includegraphics{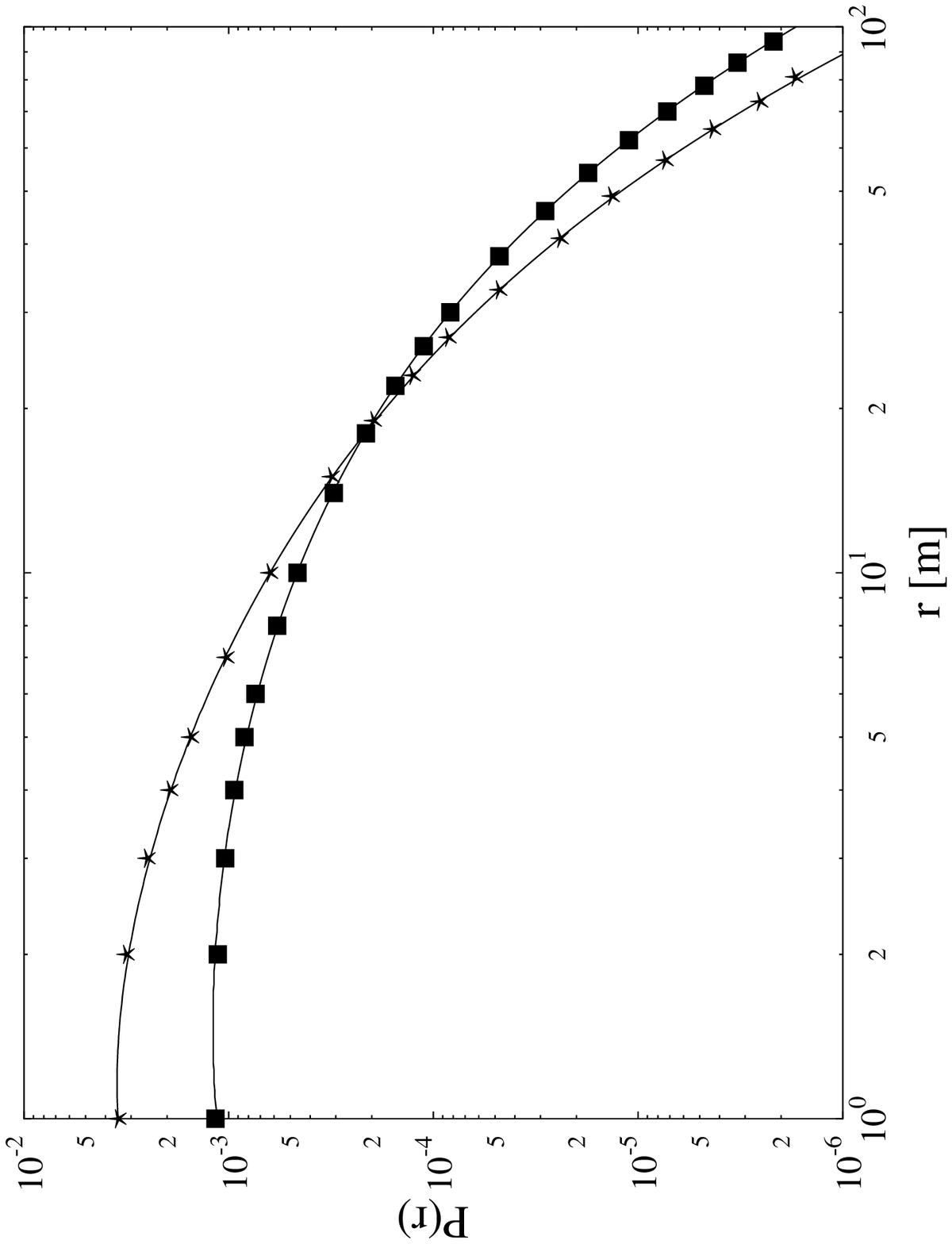}
\end{picture}
\vspace{-0.5cm}
\end{figure}
\vspace{-0.5cm}
\begin{minipage}[h]{6.8cm}
\noindent
Fig.2 Lateral distribution of muons in the bundles with multiplicities
$90 < N_{\mu} < 110$, which originated from primary proton (stars)
and primary strangelet with mass number $A=400$ (squares) (both with
energy $10^4$ TeV per particle).
\end{minipage}
\vspace{0.5cm}

The integral multiplicity distribution of muons from ALEPH data are
compared with our simulations in Fig. 1. We have used here the so
called "normal" chemical composition of primaries \cite{NIK} with
$40$ \% of protons, $20$ \% of helium, $20$ \% of C-N-O mixture, $10$
\% of Ne-S mixture and $10$ \% of Fe. It can describe low
multiplicity ($N_{\mu} \le 20$) region only. On the other hand, muon
multiplicity from strangelet induced showers are very broad. As can
be seen, the small amount of strangelets (with the smallest possible
mass number $A=400$, i.e., the one being just above the estimated
critical one estimated to be $A_{crit}\sim 320$ here) in the primary
flux can accomodate experimental data. Taking into account the
registration efficiency for different types of primaries one can
estimate the amount of strangelets in the primary cosmic flux. To
describe the observed rate of high multiplicity events one needs the
relative flux of strangelets $F_S/F_{total}\simeq 2.4\cdot 10^{-5}$
(at the same energy per particle).

\begin{figure}[h]
\setlength{\unitlength}{1cm}
\begin{picture}(7.28,7.11)
\includegraphics{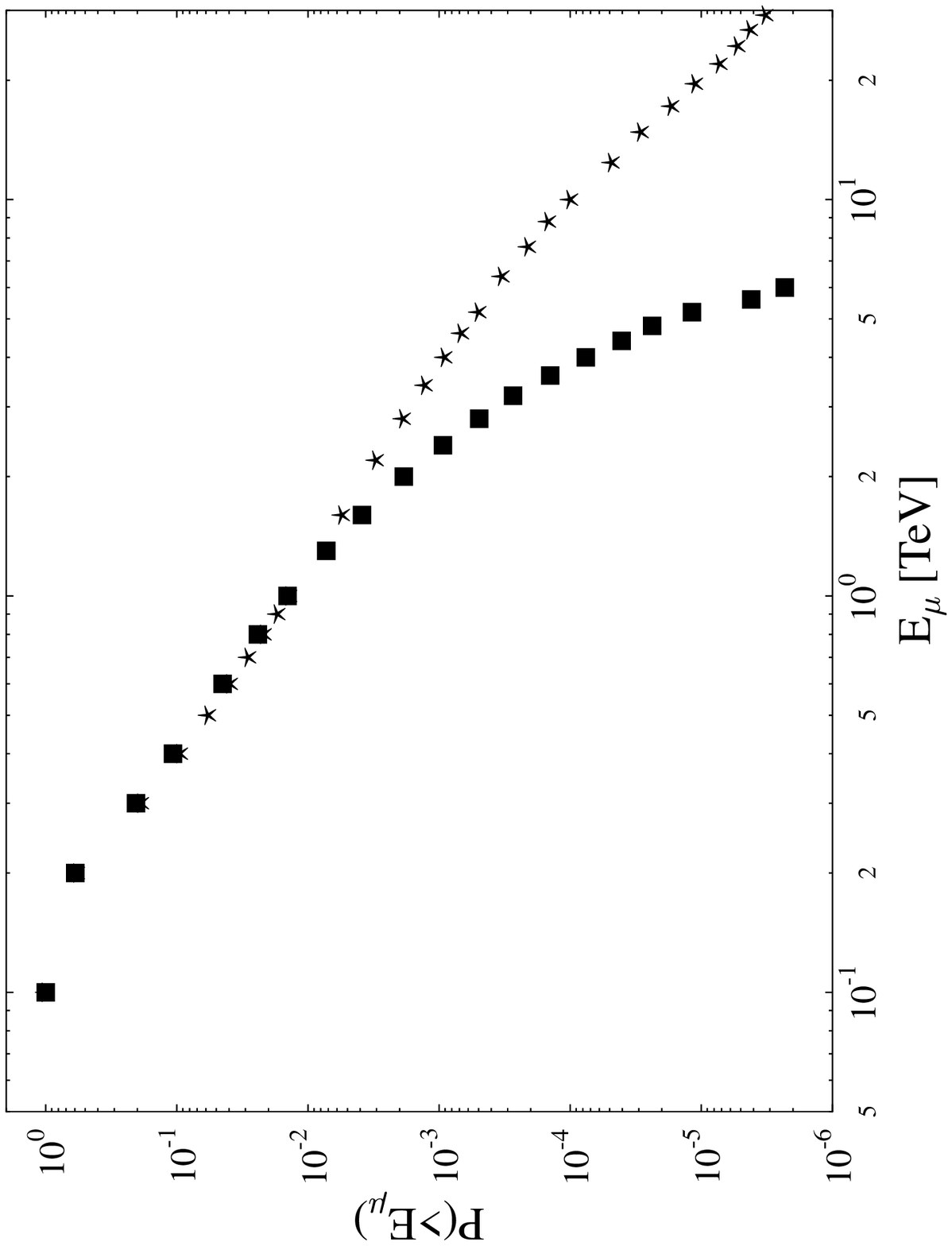}
\end{picture}
\vspace{-0.5cm}
\end{figure}
\vspace{-0.5cm}
\begin{minipage}[h]{6.8cm}
\noindent
Fig.3 Energy distribution of muons in the bundles with multiplicities
$90 < N_{\mu} < 110$, which originated from primary proton (stars)
and primary strangelet with mass number $A=400$ (squares) (both with
energy $10^4$ TeV per particle).
\end{minipage}
\vspace{0.5cm}

Fig. 2 shows the lateral distribution and Fig. 3 the energy
distribution of muons in the bundles. They allow us to test the
origin of high multiplicity events. Note that, in order to obtain
high $N_{\mu}$ tail from normal nuclei only, one needs much higher
primary energies per nucleon than in the  case where strangelets were
also added. Distribution of muons from strangelets is broader and
their energy spectrum softer in comparison to events with the same
$N_{\mu}$ induced by protons. It is interesting to observe that the
high multiplicity events discussed here (with $N_{\mu} \simeq 110$
recorded on $16~{\rm m}^2$) correspond to $\sim 5600$ muons with
$E_{\mu} \ge 70$ GeV (or $1000$ muons with energies above $220$ GeV).
These numbers are in suprisingly good agreement with results from
other experiments like Baksan Valley, where 7 events with more than
$3000$ muons of energies exceeding $220$ GeV were observed \cite{BAKSAN}.

\section{CONCLUSIONS}

To recollect: we have demonstrated that the recently observed
extremaly high multiplicity of muons can be most adequately described
by relatively minuite (of the order of $\sim 2.4\cdot 10^{-5}$ of
total primary flux) admixture of strangelets of the same total
energy. This is precisely the flux we have estimated some time ago
\cite{WW} when interpreting direct candidates for strangelets and is
fully consistent with existing experimental estimations provided by
\cite{SHF}. It accomodates also roughly the observed flux of Centauro
events as was shown in \cite{WWC}. The CosmoLEP studies of muli-muon
bundles will therefore significantly improve our understanding of the
nature and importance of the SQM candidates.\\

Acknowledgements: The partial support of Polish Committee for 
Scientific Research (grants 2P03B 011 18 and 
621/E-78/SPUB/CERN/P-03/DZ4/99) is acknowledged.\\

\end{document}